\documentclass[a4paper,twocolumn]{esapub}
\usepackage{ifluatex,ifxetex} 
\ifluatex
\usepackage{fontspec}
\else\ifxetex
\usepackage{fontspec}
\else
\usepackage{times}
\fi

\usepackage[numbers]{natbib}
\usepackage{graphicx}
\usepackage{amssymb}
\usepackage{amsmath}
\usepackage{booktabs}
\usepackage{gensymb}
\usepackage{hyperref}
\usepackage{stfloats}

\title{Simulation of Sky Surveys with the Flyeye Telescope}

\author{Olga Ram\'irez Torralba}
\affil{ESA/ESOC, Robert-Bosch-Str. 5, 64293 Darmstadt (Germany), Email: Olga.Ramirez.Torralba@esa.int\newline
TU Delft, Mekelweg 5, 2628 CC Delft (Netherlands)}
\author{R\"udiger Jehn}
\affil{ESA/ESOC, Robert-Bosch-Str. 5, 64293 Darmstadt (Germany), Email: Ruediger.Jehn@esa.int}
\author{Detlef Koschny}
\affil{ESA/ESTEC, Keplerlaan 1, 2201 AZ Noordwijk (Netherlands), Email: Detlef.Koschny@esa.int\newline
Lehrstuhl f\"ur Raumfahrttechnik, TU Munich, Boltzmannstr 15, 85748 Garching (Germany)}
\author{Michael Fr\"uhauf}
\affil{ESA/ESOC, Robert-Bosch-Str. 5, 64293 Darmstadt (Germany), Email: Michael.Fruehauf@esa.int}
\author{Laura S. Jehn}
\author{Alexander Praus}
\affil{TU Darmstadt, Karolinenplatz 5, 64289 Darmstadt (Germany)}

\begin{document}


\maketitle

\begin{abstract}
ESA's Flyeye telescope is designed with a very large field of view (FoV) in order to scan the sky for unknown near-Earth Objects (NEOs). For typical exposure times of 40\;s, the telescope is able to detect objects with a limiting magnitude up to 21.5. The aim is to observe those NEOs that are going to hit the Earth within a few weeks or days, in advance of impact.

In order to estimate the detection rate of NEOs with the Flyeye telescope, a synthetic population of Earth-threatening asteroids is created by means of the software NEOPOP. Then, the true anomalies and longitudes of the ascending node of these objects are modified in order to generate about 2\,500 impactors.

In the simulations almost three impacts can be detected per year from NEOs down to 1 m using only one Flyeye telescope. When operating two telescopes simultaneously, one at Monte Mufara and one at La Silla, four detected impacts per year are expected. Nonetheless, it is estimated that about 15.6\% of the Earth impactors will be very difficult to be detected using ground-based telescopes due to the fact that they are approaching us from the Sun. 
 
\end{abstract}

\section{Introduction}
On 15 February 2013, a small asteroid of about 20 m entered the atmosphere over Chelyabinsk and exploded about 30 km above the city. The explosion unleashed an energy equal to 20 or 30 times the energy released in the Hiroshima atomic explosion and the resulting shock wave shattered many windows, injuring nearly $1\,500$ people \cite{chelyabinsk}. Only a very small fraction of the asteroid population, in the diameter range comparable to this asteroid, has been discovered to date.

Recognising the threat of smaller near-Earth objects (NEOs), ESA's first NEO Survey Telescope, the Flyeye telescope, is aimed at detecting NEOs with apparent magnitudes up to 21.5. While American surveys, such as the Catalina Sky Survey (CSS) and the Panoramic Survey Telescope \& Rapid Response System (Pan-STARRS), are focused on finding the NEO population larger than 140 meters\footnote{\url{https://cneos.jpl.nasa.gov/stats/}}, the Flyeye telescope is mainly built to search for the smaller asteroids, \textit{i.e.}, those with less than 500 m in diameter, which are expected to collide with Earth regularly in the next decades. Under favourable conditions, the Flyeye telescope is expected to be able to detect NEOs down to 40 m in diameter, at least three weeks before impact. In this work, we estimate the performance of the Flyeye telescope in the future detection rate of the smallest NEOs in the population, with diameters down to 1 m. 

To this aim, a population of synthetic Earth impactors is created based on the validated population model by Granvik et al.\;(2018) \cite{GRANVIK2018181}. Since this population model is calibrated in the absolute magnitude range between $H=15$ and $H=25$, an extrapolation method is applied in order to extend the absolute magnitude range up to $H=32$, and thus include the small asteroids in the 1--3 m diameter range. Then, an asteroid detection software focused on the Flyeye telescope characteristics is developed, in order to check whether an asteroid can be observed before impact.

Section \ref{sec:popGen} presents the population model for the generation of synthetic NEOs and compares the population estimates to others found in literature. The extrapolation method applied to generate fainter NEOs outside the calibrated range of the model is also presented. Then, Section \ref{sec:derivImp} explains how the population of Earth impactors is derived and validated using theoretical models based on observation data. Section \ref{sec:flyeye} describes the detection conditions for NEOs and the most relevant parameters of the Flyeye telescope. Finally, Section \ref{sec:skysurvey} presents the simulation results of the sky surveys and estimates the detection rate of Earth impactors using the Flyeye telescope. 

\section{Generation of a synthetic near-Earth objects population}\label{sec:popGen}
A population of synthetic near-Earth objects (NEOs) is generated using ESA's Near-Earth Object Population Observation Program (NEOPOP), which provides the orbits and size distributions for NEOs in the absolute magnitude range between $H=15$ and $H=25$.

Since the Flyeye telescope has been designed in order to search for the small asteroids, the generated NEO population shall include those objects that are fainter than $H=25$. To this aim, an extrapolation in absolute magnitude is done in order to obtain those NEOs with $H>25$.

\subsection{The near-Earth object orbit and size population model}
The orbital and size distribution of our synthetic NEO population is derived from the validated model developed by Granvik et al.\;(2018) \cite{GRANVIK2018181}. This population model is based on a dynamical source model, which was calibrated using Catalina Sky Survey observations in the years 2006--2011. It provides debiased NEO orbit distributions of semimajor axis $a$, eccentricity $e$, inclination $i$ and absolute magnitude $H$ in the range $17<H<25$. 

The synthetic population of NEOs based on this model has been generated using the tool NEOPOP, which was developed in 2015 within the framework of activities related to ESA's Space Situational Awareness (SSA) programme\footnote{More information can be found in: \url{http://neo.ssa.esa.int/neo-population}}. Although the population model is calibrated in the interval $17<H<25$, the software has implemented an extrapolation method in order to extend it to $H=15$, and a database of known NEOs is used at the bright end $H<15$ since the NEO population in that range is assumed to be completely known. 

\subsection{Extrapolation to smaller or fainter near-Earth objects}\label{sec:extrapolmodel}
As pointed out by Granvik et al.\;(2018) \cite{GRANVIK2018181}, the limitation of extrapolating the population model to smaller (and/or larger) objects is that the orbit distributions are fixed to either $H=25$ or $H=17$, respectively. In order to obtain orbit distributions for the smallest NEOs with absolute magnitudes above $H=25$, the extrapolation method used by the Population Generation tool of NEOPOP is as follows. 

When extrapolating to smaller NEOs, NEOPOP uses the following power-law function to evaluate the total number of objects when the required absolute magnitude $H$ exceeds the calibrated range of the model ($H>25$):
\begin{equation}
    \textrm{pobj} = \textrm{totp}\left(10^{(\alpha_{\textrm{extr}}(H-25))}-1\right)
\end{equation}
where $\textrm{pobj}$ is the predicted number of objects from $H=25$ to the required $H$, $\textrm{totp}=802404$ is the cumulative number of objects for the full model ($15 < H < 25$) and $\alpha_{\textrm{extr}}$ is the slope of the power law. This parameter is set to a fixed value of 0.6434, which was estimated by Brown et al.\;(2002) \cite{brown2002flux} from satellite observations of bolides and resulted in an extrapolation that is in good agreement with the literature.

Unfortunately, the Population Generation tool of NEOPOP does not allow the creation of NEOs with $H>30$ due to the fact that the extrapolation method used is very imprecise for such small objects. It is not possible to find NEO orbit models in literature for NEOs with $H>30$.

The main reason is that population and distribution models of the absolute magnitudes for near-Earth objects are derived from the set of all currently known NEOs. However, the large fraction of the NEO population and especially those with $H>29$ ($\lesssim$ 5 m in diameter) remain undetected. 

Therefore, if we are interested in those NEOs in the $1--3\;\mathrm{m}$ diameter range, another extrapolation approach is needed. This is discussed in Section \ref{sec:genImp}.

\subsection{Comparison with other population estimates}\label{sec:popComparison}
For a population of NEOs with absolute magnitudes between $H=17$ and $H=30$, the number of objects predicted by both the model from Granvik et al.\;(2018) \cite{GRANVIK2018181} and the extrapolated model described in Section \ref{sec:extrapolmodel} are presented in Table \ref{tab:numberNEOs}.

The total number of NEOs in the desired $H$ range is estimated to be roughly $1.32\cdot10^9$. However, since the Population Generation tool has an upper limit of $10^7$ orbits, a sampling factor of 1\,000 is applied. Therefore, the sample population of synthetic NEOs that is used in this work has a total of $1.32\cdot10^6$ objects. Due to this sampling factor that must be applied, each one of the NEOs in the sample population is assumed to be representative of $1\,000$ similar NEOs.

\begin{table}[]
\centering
\begin{tabular}{@{}ccc@{}}
\toprule
 & $H$ & Total number of NEOs \\ \midrule
Granvik \cite{GRANVIK2018181} & $[17.00, 25.00]$ & $800\,000$ $\pm$ $100\,000$ \\ \midrule
Extrapolated & $[17.00, 30.00]$ & 1.3$\cdot10^9$ $\pm$ 1.74$\cdot10^8$ \\ \bottomrule
\end{tabular}
\caption{Predicted number of NEOs between $H=17$ and $H=30$.}
\label{tab:numberNEOs}
\end{table}

The resulting NEO population is compared to the population estimates obtained by Stokes et al.\;(2017) \cite{stokes2017update}, who uses an orbital distribution derived from the same model by Granvik et al.\;(2018) \cite{GRANVIK2018181} but the absolute magnitude and size distributions adopted are different.

\begin{figure}
    \centering
    \includegraphics[width=1.0 \columnwidth]{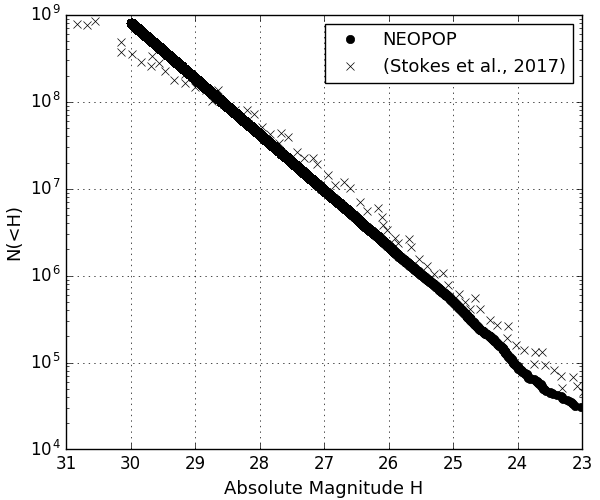}
    \caption{Comparison of cumulative NEOs distribution between the extrapolated model in NEOPOP based on \cite{NEOPOP} \rm{\footnotesize{($\bullet$)}} {\it and the population estimates extracted from \citep{stokes2017update}} \rm{\footnotesize{($\times$)}}.}
    \label{fig:comparisonNASAvsESA}
\end{figure}

Figure \ref{fig:comparisonNASAvsESA} compares the cumulative NEO distribution obtained using the Population Generation tool of NEOPOP and the results derived in \citep{stokes2017update}. Although there is a relatively good agreement between both population estimates, it is possible to see that, in general, the extrapolated model in NEOPOP that is used in this work predicts a lower number of NEOs, except in the range $H>29$ for which the number of smaller NEOs is overestimated.

\section{Derivation of a population of synthetic Earth impactors}\label{sec:derivImp}
Once the sample population of NEOs is generated, a population of Earth impactors can be derived. The mean anomaly and longitude of the ascending node of appropriate objects are modified to obtain as many colliding objects as possible in order to analyse the performance of the Flyeye telescope in a statistically meaningful manner.

\subsection{Close approach analysis and impact detection}
The population of synthetic Earth impactors is derived by finding those NEOs whose distance of closest approach (DCA) to Earth is lower than the Earth's radius. To this aim, each object of the population is propagated for one year and a close approach analysis is performed in order to determine the minimum distance to Earth.

Due to the high computational effort required, a simple Kepler propagation of the heliocentric orbit of each NEO is used. In order to consider the influence of the gravitational attraction of Earth, it is assumed that an impact happens when the DCA is lower than the effective radius of the Earth $b_{\oplus}$, which includes the gravitational focusing as described by \cite{valsecchi2003resonant}, and is estimated with:
\begin{equation}
    b_{\oplus}=R_{\oplus}\sqrt{1+\frac{2\mu_{\oplus}}{R_{\oplus}v_{\infty}^2}}
\end{equation}
where $R_{\oplus}$ is the radius of Earth, $\mu_{\oplus}$ is the gravitational parameter of Earth and $v_{\infty}$ is the unperturbed encounter velocity of the NEO with respect to the Earth. 

A typical value for $v_{\infty}$ for impacting NEOs can be taken from the work by \cite{valsecchi2003resonant}. Assuming a mean value of roughly $v_{\infty}\approx12$ km/s leads to an effective radius of 1.37 times the Earth's radius.

For the close approach analysis, the DCA is determined by using the Differential Evolution optimisation algorithm \cite{DEalgorithm}. This optimisation method is a stochastic population-based method that is useful for global optimisation problems when the search space is very large. Out of all the different optimisation algorithms that are provided in the Python library SciPy, this proved to be the most efficient.

\subsection{Generation of Earth impactors}\label{sec:genImp}
The NEOs are assumed to be on unperturbed Keplerian orbits, which are described by the orbital elements $(a,e,i,\Omega_{0},\omega,M_{0})$, where $\Omega_{0}$ is the longitude of the ascending node, $\omega$ is the argument of periapsis and $M_{0}$ is the mean anomaly at the reference epoch $t_{0}$. 

Although our synthetic population consists of 1.3 million objects, the probability that any of them hits the Earth during a simulation period of one year is low and in any case not sufficient for any statistical analysis. However, since the longitude of the ascending node and the mean anomaly were randomly selected, we can modify them in order to get a sufficiently large number of impactors.

We apply the Differential Evolution algorithm to find the minimum distance of closest approach with three independent variables: the time, the mean anomaly at the reference epoch and the longitude of the ascending node of the orbit. The upper and lower bounds of these variables are set to:
\begin{align*}
    t &\in [t_{0}, t_{0} + 1yr] \\
    M &\in [0, 2\pi] \\
    \Omega &\in [\Omega_{0} - 5\degree, \Omega_{0} + 5\degree]
\end{align*}

Within these intervals a total of 2\,451 objects are found to have a DCA lower than the effective radius of Earth. For our simulated population, they are all in an interval of absolute magnitudes between $H=23$ and $H=30$.

Still, we are interested in simulating even smaller objects, in order to better estimate the Flyeye telescope's detection rate of the smallest NEOs. To achieve this we apply a little trick: the absolute magnitude of each object is increased by two units, \textit{i.e.}, objects that were created with an absolute magnitude of 29 are now assigned a magnitude of 31. We are assuming that the orbital distribution of the objects hardly changes for such a small change in magnitude. 

This means that the impactor population is now defined in the interval $25<H<32$, which corresponds to a diameter range of 1--34 m. Moreover, due to this shift in $H$, the sampling factors of the Earth impactors must be adapted since more objects are expected. 

\subsection{Adaptation of the sampling factor}
In order to determine the slope of increase in impacting NEOs as function of size or absolute magnitude, use is made of the power-law fit to observation data derived by Brown et al.\;(2002) \cite{brown2002flux}. The cumulative number of objects colliding with the Earth each year $N$ with diameters of $D$ (in meters) or greater is described by:
\begin{equation}
    \log_{10}{N} = c_0 - d_0\log_{10}{D}
    \label{eq:CumNumbervsD}
\end{equation}
where $c_0=1.568\pm0.03$ and $d_0=2.70\pm0.08$.

Since the NEO population is described in terms of absolute magnitude $H$ and not size, the approximate size of an object is derived from $H$ and the albedo (assumed to be $p=0.15$) using the following equation:
\begin{equation}
    D = 1.329\cdot10^6p^{-0.5}10^{-0.2H}
    \label{eq:DvsH}
\end{equation}

The required scaling factor due to the shift in magnitude can be easily computed by means of Eqs. \ref{eq:CumNumbervsD} and \ref{eq:DvsH}. When increasing $H$ by two, the cumulative number of Earth impactors per year is increased by a factor of approximately 12.

\begin{table}[]
\centering
\begin{tabular}{@{}cc@{}}
\toprule
$H$ & Sampling factor $k_{\mathrm{samp}}$ \\ \midrule
$(25.0, 30.0]$ & 24\,000 \\
$(30.0, 32.0]$ & 12\,000 \\ \bottomrule
\end{tabular}
\caption{Sampling factors for different $H$ ranges.}
\label{tab:fittingfactors}
\end{table}

Moreover, in Figure \ref{fig:comparisonNASAvsESA} it was observed that our population is by a factor of about 2 too small for $H < 28$. Therefore, the sampling factor is doubled in this region. Together with the sampling factor of 1\,000 introduced in Section \ref{sec:popComparison}, the overall sampling factor is now 12\,000 and 24\,000, respectively. Table \ref{tab:fittingfactors} summarises the sampling factors that are applied to our impactor population for the subsequent simulations.

\subsection{Cumulative number of Earth impactors per year}\label{sec:cumulativenumberimpactors}
Since our 2\,451 Earth impactors have been generated by modifying their mean anomalies $M$ and longitudes of the ascending node $\Omega$, we need to calculate the probability of the possible $M$ and $\Omega$ combinations for which the NEO is impacting the Earth, assuming that the mean anomaly is uniformly distributed between 0 and $2\pi$ and $\Omega$ is uniformly distributed between $\Omega_{0} - 5\degree$ and $\Omega_{0} + 5\degree$.

The approach to estimate this probability is as follows. First, we compute the interval of longitude of the ascending node $[\Omega_{0},\Omega_{f}]$, where we have an impact. Then, we divide this interval in $\textrm{N}$ bins (we chose $\textrm{N}=10$) with widths $\Delta{\Omega}$ and, for each value $\Omega_{i}$ in the interval, we estimate the probability of the object to have the `right' mean anomaly in order to impact the Earth:
\begin{equation}
    P_{M}(\Omega_{i}) = \frac{M_{f}(\Omega_{i})-M_{i}(\Omega_{i})}{2\pi}
\end{equation}
where $[M_{0}(\Omega_{i}),M_{f}(\Omega_{i})]$ is the interval of mean anomaly, for a fixed $\Omega_{i}$, such that we have an impact.

\begin{figure}
    \centering
    \includegraphics[width=1.0 \columnwidth]{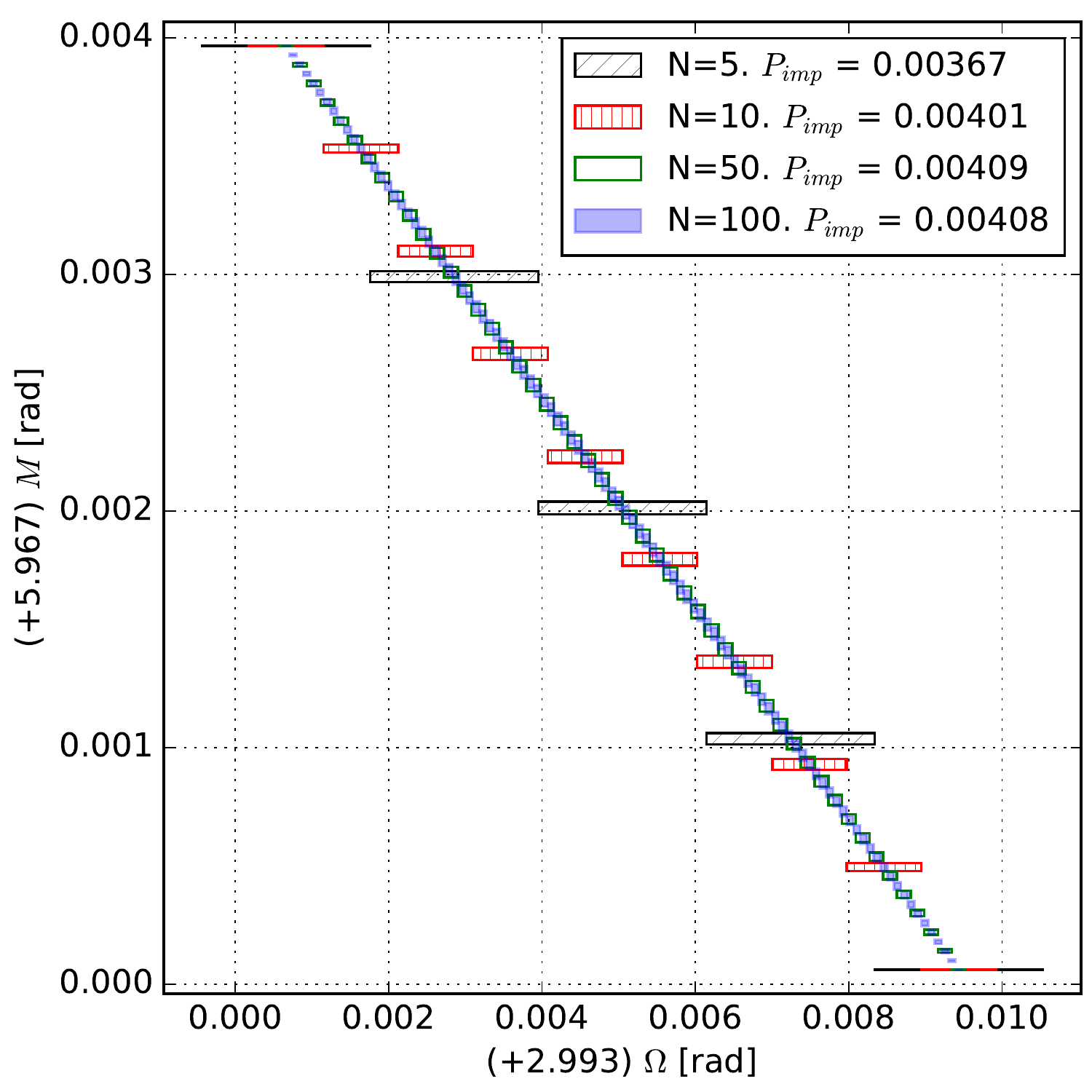}
    \caption{Impact area in the mean anomaly $M$ -- longitude of ascending node $\Omega$ space. The impact probability is proportional to the size of the area. A bin size of $\textrm{N}=10$ in $\Omega$ is sufficient to estimate the size of the area.}
    \label{fig:areaprob}
\end{figure}

Finally, the probability of each impactor to impact the Earth in one year is computed with:
\begin{equation}
    P_{\textrm{imp}} = \sum_{i}^\textrm{N}P_{M}(\Omega_{i})\frac{\Delta{\Omega}}{10\degree}k_{\mathrm{samp}}
\end{equation}

Figure \ref{fig:areaprob} shows the impact area in $M$ and $\Omega$ for a random impactor in the population, obtained using different number of bins $\textrm{N}$. First, it is possible to observe that the intervals of both $M$ and $\Omega$, where we have an impact, are extremely narrow. Second, although using $\textrm{N}=10$ seems to result in a very crude approximation when compared to $\textrm{N}=100$, the obtained probabilities of impact $P_{imp}$ are very similar. Therefore, since it was very computationally consuming to do the computations with a lower resolution, $\textrm{N}=10$ was selected. 

When adding up the probabilities of all impactors, 17.4 objects with absolute magnitudes lower than $H=32$, \textit{i.e.}, with diameters larger than about 1 m (assuming an albedo of 0.15), are predicted to collide with Earth in one year. 

\begin{figure}
    \centering
    \includegraphics[width=1.0 \columnwidth]{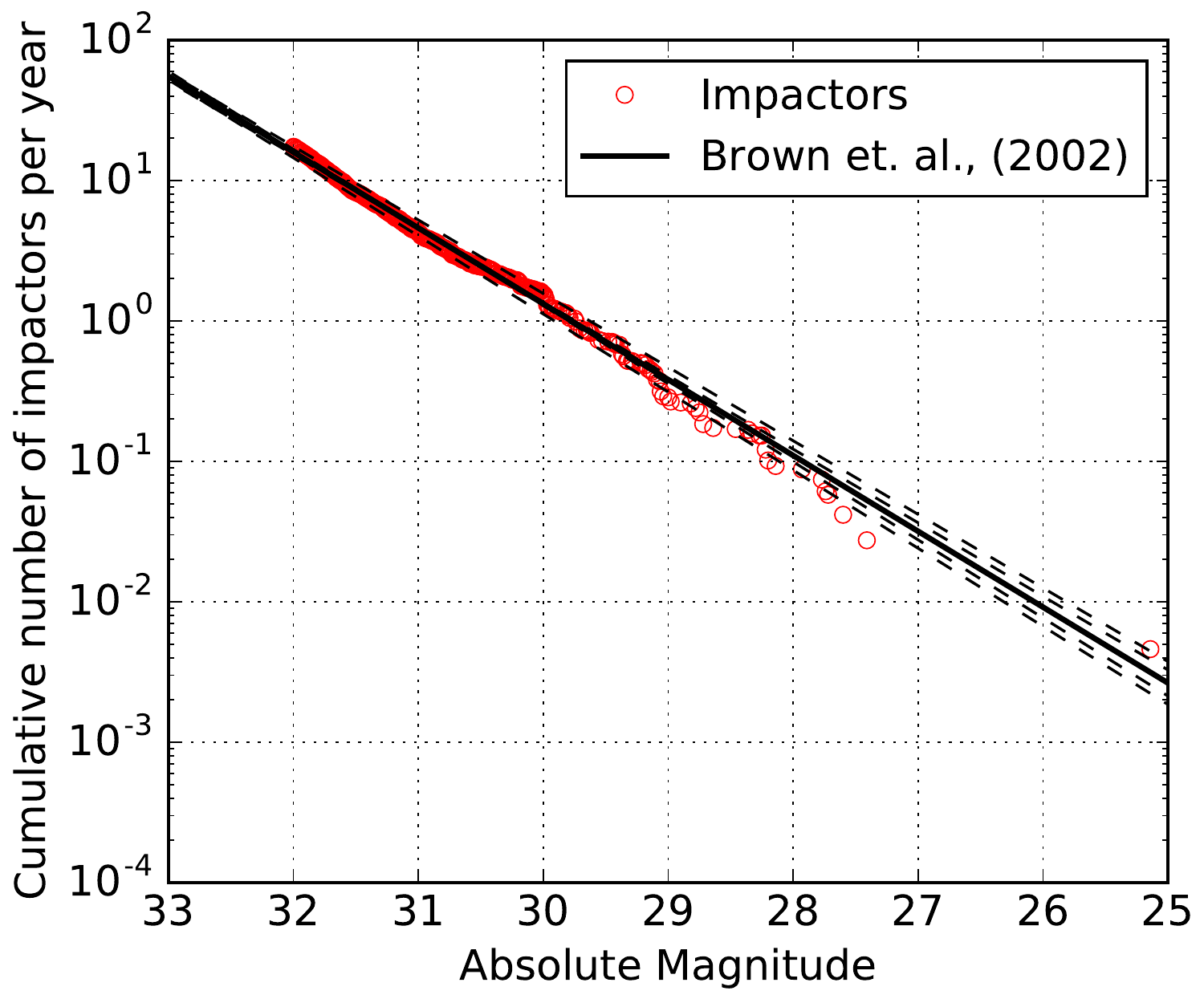}
    \caption{Comparison in the estimated cumulative number of objects colliding with the Earth between our impactor population model \rm{\footnotesize{($\circ$)}} {\it and the theoretical model by Brown et al.\;(2002) \cite{brown2002flux}} \rm{\footnotesize{(\textemdash)}}. {\it The dashed lines represent the uncertainties in the theoretical model.}}
    \label{fig:calibratedCumNumImpactors}
\end{figure}

Figure \ref{fig:calibratedCumNumImpactors} shows the cumulative number of Earth impactors derived by summing up the $P_{imp}$ of all 2\,451 impactors, and compares it to the theoretical model by Brown et al.\;(2002) \cite{brown2002flux}. Our estimates are in very good agreement with the model, which indicates that the probability of impact for each NEO was correctly computed. 

\section{Observation of near-Earth objects using the Flyeye Telescope}\label{sec:flyeye}
This section describes how the detectability of NEOs is determined. The most relevant parameters of the Flyeye telescope are also introduced. 

\subsection{Required detection conditions for near-Earth objects}\label{sec:detectcond}
The first requirement to be able to detect an object is that the Sun's elevation is below the horizon, although detections are unlikely for an elevation above -14$\degree$. For those objects arriving during twilight, the last condition regarding the Signal-to-Noise Ratio (SNR) will filter out those that are too faint. 

Second, due to the mechanical structure of the telescope, the asteroid's elevation must be above 15$\degree$.

Then, for a nominal exposure time of 40 s, the Flyeye telescope is able to observe objects up to a magnitude of 21.5. Thus, the apparent magnitude of the asteroid must be below this limiting magnitude in order to be able to be detected by the telescope. The computation of the apparent magnitude of the NEO is discussed in Section \ref{sec:appmag}.

The last condition is regarding the SNR of the object, which compares the level of a desired signal to the level of the noise, and determines whether an object is detectable in an image. Based on experience, an SNR $> 5$ is considered a robust criterion for a reliable detection. The method implemented for the computation of the SNR is explained in Section \ref{sec:SNR}. 

To sum up, an asteroid can be detected if:
\begin{itemize}
    \item The apparent magnitude of the object is below 21.5.
    \item The asteroid has an elevation above 15$\degree$.
    \item The Sun's elevation is below 0$\degree$.
    \item The SNR is above 5.
\end{itemize}

\subsection{Apparent magnitude of a near-Earth object}\label{sec:appmag}
In the NEO field, the most widely used method to compute the apparent magnitude of an object is the one recommended by the International Astronomical Union (IAU), as defined in \cite{appmagnitude}. However, this method is not valid for phase angles greater than 120$\degree$ and is best used at much smaller values, 20$\degree$ or less, \textit{i.e.}, for main-belt asteroids.

\begin{figure}
    \centering
    \includegraphics[width=1.0 \columnwidth]{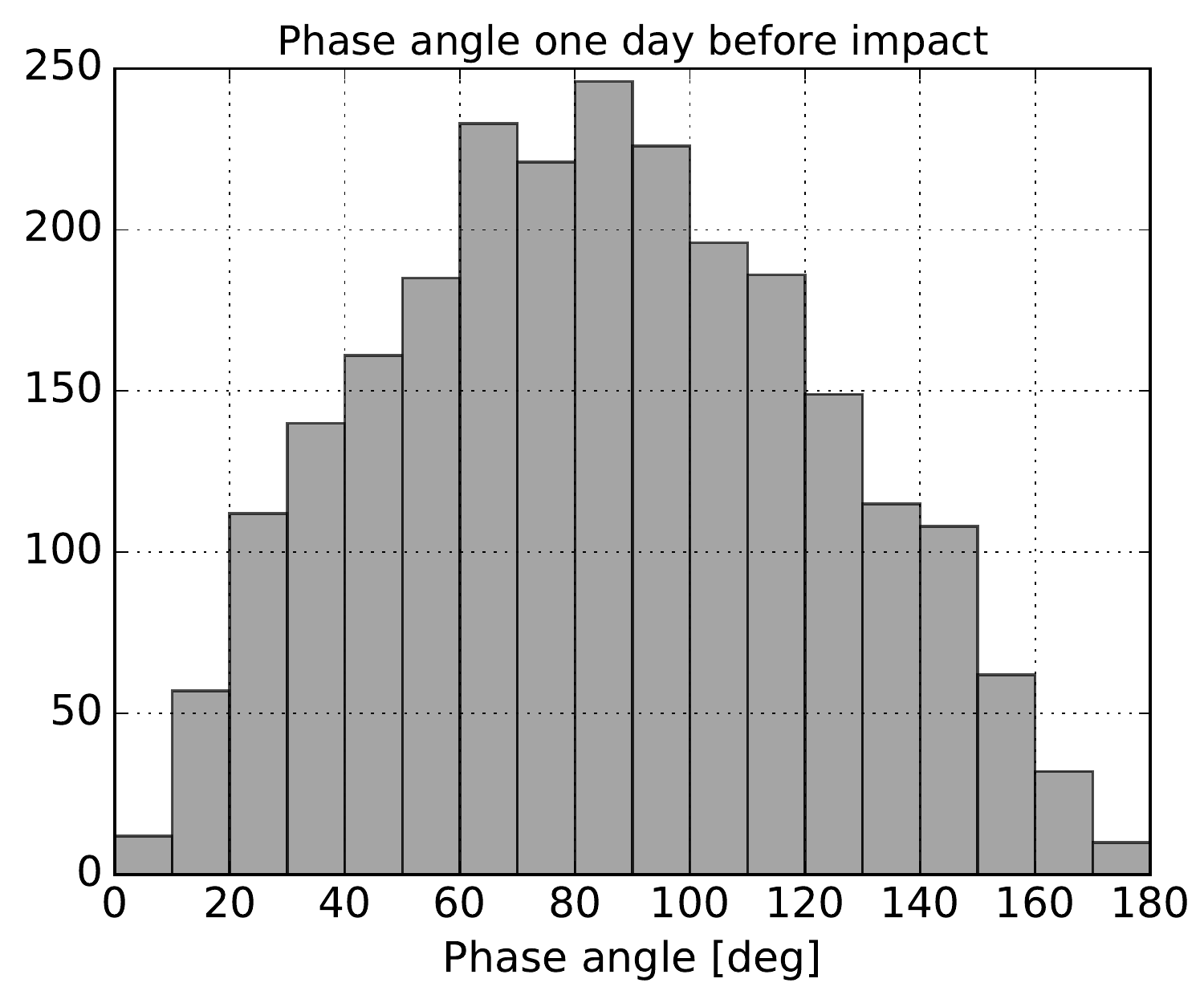}
    \caption{Histogram plot of the phase angles of all impactors one day before impact.}
    \label{fig:pha1daybefimp}
\end{figure}

As Figure \ref{fig:pha1daybefimp} shows, there are many objects in our population that are approaching with a high phase angle, larger than 120\degree, and thus the IAU model cannot be used. 

For this reason, the model defined in \cite{NEOPOP} is used, which does not have any restriction on the phase angle. In this model, the apparent magnitude of the NEO $m_{\mathrm{ast}}$ is computed using an approach for solar system bodies:
\begin{equation}
    m_{\mathrm{ast}} = H\left(1,\alpha\right) + 2.5\log_{10}\left(\frac{r^2\Delta^2}{\phi(\alpha)}\right)
\end{equation}
where $r$ is the distance of the NEO from the Sun (in au), $\Delta$ is the distance of the NEO from the Earth (in au), $\alpha$ is the phase angle, \textit{i.e.}, the angle Sun-object-observer, and $\phi(\alpha)$ is the phase function:
\begin{equation}
    \phi(\alpha) = \frac{2}{3}\left(\left(1-\frac{\alpha}{\pi}\right)\cos{\alpha}+\frac{1}{\pi}\sin{\alpha}\right)
\end{equation}

The opposition effect is described by computing the absolute magnitude of the NEO $H$ as a function of the phase angle:
\begin{equation}
    H\left(1,\alpha\right) = H - \frac{a}{1+\alpha} + b\cdot\alpha
\end{equation}
where the values $a$ and $b$ depend on the albedo. For an assumed albedo of $p=0.15$, $a=0.340$ and $b$ is computed according to the phase angle $\alpha$:

\begin{align}
    & b = 0.0125 - 0.01045\cdot\ln{p}\qquad\mathrm{for}\quad\alpha < 5\degree\\
    & b = 0.013 - 0.024\cdot\log_{10}{p}\,\qquad\mathrm{for}\quad\alpha > 5\degree
\end{align}

\subsection{Signal-to-Noise Ratio}\label{sec:SNR}
There are many different ways to calculate the SNR of an object and there is still an open discussion regarding which is the best method. In this work, the method defined by \cite{Detlef} is implemented, which only looks at the SNR in the centre pixel. 

The most rigorous way, but also more complicated, would be to consider the spread of the light source over several pixels and integrate it to obtain the volume of the light distribution. However, due to the Gaussian shape of the light source, it can be assumed that most of the light is collected in the centre pixel, and thus only the percentage of light within this centre pixel is considered.

The SNR is computed with the following equation:
\begin{equation}
\resizebox{1.0\columnwidth}{!}{$\textrm{SNR} = \frac{\textrm{DN}_{\mathrm{signal}}}{\left({\textrm{DN}_{\mathrm{signal}} + \textrm{DN}_{\mathrm{bias}} + \textrm{DN}_{\mathrm{dark}} + \textrm{DN}_{\mathrm{readout}} + \textrm{DN}_{\mathrm{sky}}}\right)^{1/2}}$}
\end{equation}
where $\textrm{DN}_{\mathrm{signal}}$ is the Digital Number (DN) of the arriving signal and the rest of the terms describe the different sources of noise in the images.

The DN is a brightness value that relates to the number of electrons stored in each pixel, which are generated by the photons captured in the sensor of the CCD cameras. 

The DN of the arriving signal of the object can be computed by means of the input flux and the detector properties:
\begin{equation}
    \textrm{DN}_{\mathrm{signal}} \approx t_{\mathrm{exp}}p_{\mathrm{px}}\frac{1}{g}\frac{F_{\mathrm{detect}}}{E_{\mathrm{phot}}}\textrm{\textrm{QE}}
\end{equation}
where $t_{\mathrm{exp}}$ is the exposure time, $p_{\mathrm{px}}$ is the percentage of light within the centre pixel, $g$ is the gain (number of e- per DN), $F_{\mathrm{detect}}$ is the flux at the detector, $E_{\mathrm{phot}}$ is the energy of one photon and $\textrm{QE}$ is the quantum efficiency. A value of $p_{\mathrm{px}}=0.4$ is considered as it proved to be in good agreement with observations. 

The flux at the detector can be easily computed if the optical characteristics of the telescope are known:
\begin{equation}
    F_{\mathrm{detect}} = F_{\mathrm{in}}\pi \left(\frac{D_{\mathrm{aperture}}}{2}\right)^2(1-\textrm{obstr})\tau
\end{equation}
where $F_{\mathrm{in}}$ is the incoming flux density from the object, $D_{\mathrm{aperture}}$ is the diameter of aperture, $\textrm{obstr}$ is the obstruction and $\tau$ is the optical throughput. 

The sources of noise in the images considered are the readout noise $\textrm{DN}_{\mathrm{readout}}$, the electronic or bias noise $\textrm{DN}_{\mathrm{bias}}$, the dark current noise $\textrm{DN}_{\mathrm{dark}}$ and the sky background $\textrm{DN}_{\mathrm{sky}}$. The DN of the sky brightness takes into account the apparent size of a pixel of the telescope and can be computed with:
\begin{equation}
    \textrm{DN}_{\mathrm{sky}} \approx t_{\mathrm{exp}}\textrm{pixelscale}^2\frac{1}{g}\frac{F_{\mathrm{detect}}}{E_{\mathrm{phot}}}\textrm{QE}
\end{equation}
where the pixel scale is a measure of the image resolution of the telescope. 

The incoming flux at the detector $F_{\mathrm{in}}$ coming from either the object or the sky brightness can be easily computed using the following conversion from magnitude $m$ to flux $F$:
\begin{equation}
    F = 10^{0.4(m_{\odot} - m)}F_{\odot}
\end{equation}
where $m_{\odot} = -27.1$ is the magnitude of the Sun in the standard V-band filter and $F_{\odot} = 1362$ W/m$^2$ is the solar constant.

The background magnitude (or sky brightness) is computed in terms of mag/arcsec$^2$ as a sum of different light contributions. The approach followed is not explained in detail here since it is implemented in a similar way than in NEOPOP \citep{NEOPOP}. The light sources considered are atmospherically scattered light, scattered sunlight, scattered moonlight and zodiacal light. The contribution of planets, stars, galaxies, airglow, interstellar medium and light pollution is neglected. 

\subsection{Relevant parameters and sky survey strategy of the Flyeye Telescope}
Table \ref{tab:flyEye1} summarises the optical characteristics of the telescope and Table \ref{tab:flyeye2} summarises the characteristics of the CCD camera. More information regarding the architecture of the telescope can be found in \cite{flyeye}. 

\begin{table}[]
\centering
\begin{tabular}{@{}ll@{}}
\toprule
Effective focal length                   & 2000 mm                                               \\ \midrule
Aperture diameter                        & 1180 mm                                               \\ \midrule
Obstruction                              & 26\%                                                  \\ \midrule
Optical throughput                       & 80\%                                                  \\ \midrule
Limiting magnitude at 40 s exposure time & 21.5 Vmag                                              \\ \midrule
FoV                                      & 6.7$\degree\,\times$ 6.7$\degree$ \\ \bottomrule
\end{tabular}
\caption{Flyeye telescope optical characteristics. Source:\;\cite{flyeye}}
\label{tab:flyEye1}
\end{table}

\begin{table}[]
\centering
\begin{tabular}{@{}ll@{}}
\toprule
Pixel size         & 15 $\mu$m $\times$ 15$\mu$m \\ \midrule
Pixel scale        & 1.5$''$                      \\ \midrule
Exposure time      & 40 s                       \\ \midrule
Readout noise      & 10 e-                      \\ \midrule
Quantum Efficiency & 75\%                       \\ \midrule
Gain               & 90\%                       \\ \midrule
Dark current       & 0                          \\ \midrule
Bias               & 3000 DN                    \\ \bottomrule
\end{tabular}
\caption{Flyeye CCD camera characteristics. Source:\;\cite{flyeye}}
\label{tab:flyeye2}
\end{table}

The Flyeye telescope will apply the `Wide Survey Strategy'. This observation strategy allows to scan half of the visible sky per night from one location, assuming that each field in the sky shall be covered four times to allow the asteroids to move at least a few pixels.

This sky survey strategy is not simulated in this work. The focus is to determine whether the detection conditions are satisfied for each impactor, and to obtain good statistics regarding the number of observed and missed impactors. Therefore, it is assumed that if an asteroid has an SNR high enough to be detected and is located above the horizon, the telescope will be able to observe it.

One of the key parameters that will be analysed is the warning time for each impactor. For warning times longer than two days, it is reasonable to assume that the telescope will eventually point to the right direction, and observe the asteroid before impact. However, for warning times lower than one day, there is less than 50\% chance that the telescope is not pointing at the right field of sky, and thus the asteroid is missed.

\section{Simulation of Sky Surveys}\label{sec:skysurvey}
In this section we present the average warning times and the number of impactors that will be detected at a given telescope. 

The benefit of adding a second telescope is also studied by comparing the detection rates using two different locations in the two hemispheres: Monte Mufara (Sicily, Italy) in the Northern hemisphere, and La Silla (Chile) in the Southern hemisphere.

Then, for those missed impactors, we estimate how many of them remain undetected because they are coming from the dayside.

\subsection{Observation of Earth impactor population as a function of H}\label{sec:obspopH}
The first simulations are performed assuming the Flyeye telescope located on Monte Mufara, which is where the first one will be deployed. 

Table \ref{tab:obsvsmissed} presents the relative number of observable impactors, for different ranges in absolute magnitude. The computation of the detection rates presented does not take into account the probability of impact $P_{\textrm{imp}}$ of each object, because it was verified that the results changed only slightly when weighting the detection rates with $P_{\textrm{imp}}$.

\begin{table}[]
\centering
\begin{tabular}{@{}ccc@{}}
\toprule
$H$ & Observable & Average warning time \\ \midrule
$[25, 32]$ & 47.0 \% & 0.67 days \\ \midrule
$[25, 30]$ & 61.3 \% & 1.69 days \\
$(30, 32]$ & 45.7 \% & 0.55 days \\ \bottomrule
\end{tabular}
\caption{Relative number of observable impactors and average warning time, for different ranges in absolute magnitude $H$, using the Flyeye telescope at Monte Mufara.}
\label{tab:obsvsmissed}
\end{table}

First, for the brightest range up to $H=30$, about 61\% of the objects are observed. Then, as the objects get fainter, the number of missed NEOs increases. For those objects in the diameter range between 1 and 3 m, $H>30$, about 54\% are missed.

Table \ref{tab:obsvsmissed} also presents the average warning time, \textit{i.e.}, the time between impact and observation. As expected the average warning time is very short. Moreover, it is very likely that NEOs are missed as their warning times are less than two days; this means that the visible sky is not scanned entirely and it might happen that the telescope is not pointing at the right direction to observe the asteroid. Therefore, the number of observable objects as reported in Table \ref{tab:obsvsmissed} is too optimistic.

\begin{figure}
    \centering
    \includegraphics[width=1.0 \columnwidth]{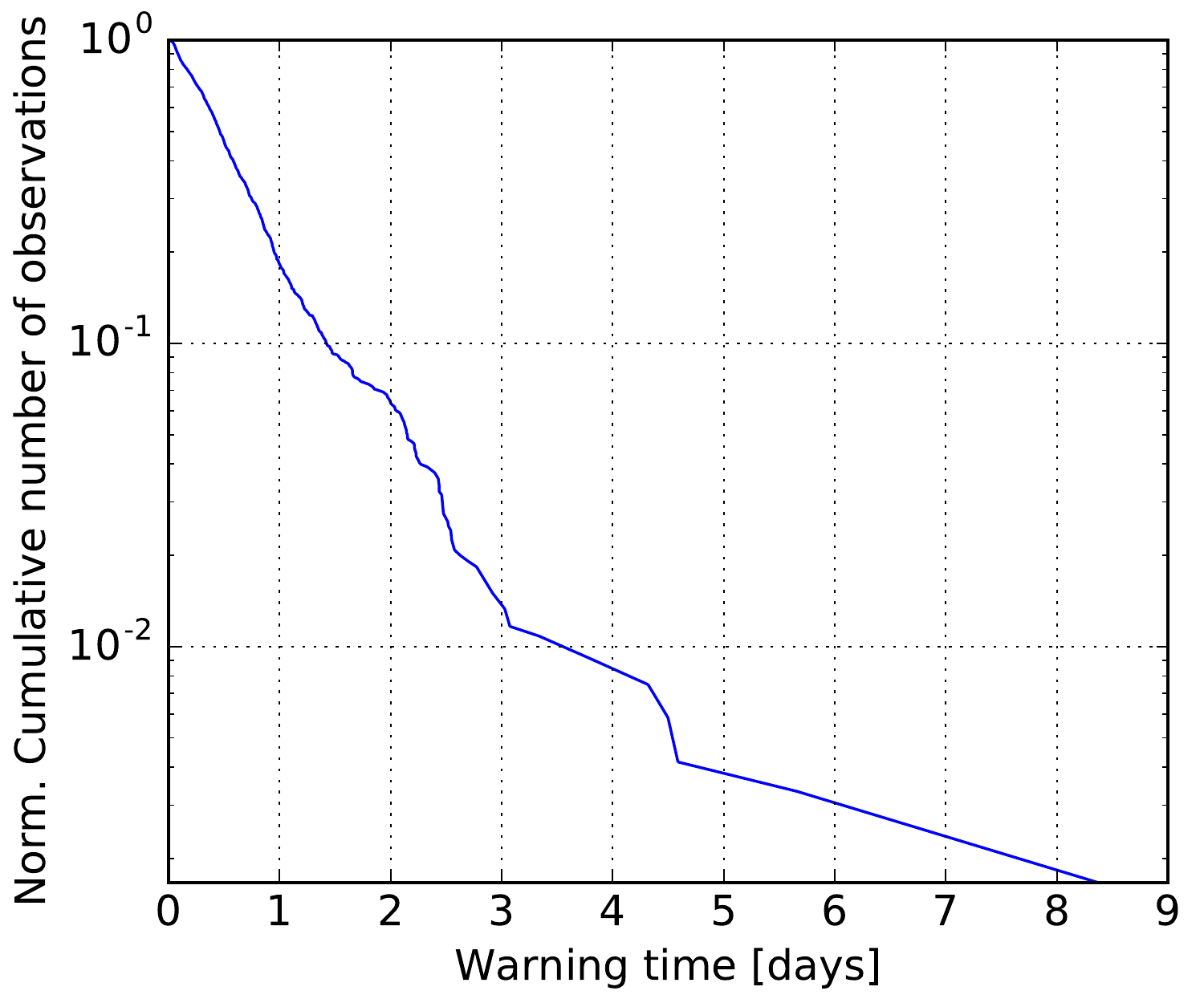}
    \caption{Cumulative number of observable impacting NEOs as a function of the warning time.}
    \label{fig:cumnumDetections}
\end{figure}

Figure \ref{fig:cumnumDetections} shows the normalised cumulative number of observable impacting NEOs as a function of their warning times. It is estimated that only about 6.4\% will have a warning time of at least two days, and thus they are likely to be detected.

The main reason for such short warning times is that we consider objects down to 1 m in diameter. The time when the apparent magnitude of the asteroid drops below the limiting magnitude of 21.5 and at the same time meets the requirement to be 15$\degree$ above the horizon of the telescope, is usually very close to the impact time. Therefore, the observation window is reduced significantly for large $H$, resulting in very short warning times.

Second, it has been observed that the observation window, \textit{i.e.}, the time between when the apparent magnitude crosses the 21.5 threshold and impact, for the brightest object of the population ($H=25.14$) is about 7.23 days, while the longest observation window (about 8.59 days) happens for an object with $H=27.41$ because of a much smaller phase angle.

Micheli et al.\;(2018) \cite{micheli} studied the average angular velocity of impactors before collision and noticed that the objects are unusually slow up to about a day before impact, with velocities below the detection limit. In reality the software to detect moving objects might not flag them because the apparent motion is below a given threshold. This effect is not considered in our simulations; this could be a new feature to be implemented in the software for future work. 

\subsection{Benefit of a second telescope in the Southern hemisphere}\label{sec:telloc}
Due to the very short warning times obtained, many impactors will be missed by a single telescope just because the impact takes places on the other side of the world. 

Here we analyse whether a second telescope located on the other hemisphere is able to observe these impactors. To this aim, the simulations are repeated but considering that a second Flyeye is located on La Silla, Chile.

Table \ref{tab:obsvsmissed2tel} presents the relative number of observable impactors when using simultaneously two Flyeye telescopes at two different locations. As expected, the number of observable NEOs has increased significantly; with two telescopes, we are able to observe 73.5\% of the entire impactor population instead of only 47\%. Moreover, about 90\% of the objects in the brightest range up to $H=30$ can be observed, which is significantly higher than the 61\% when using one telescope.

\begin{table}[]
\centering
\begin{tabular}{@{}ccc@{}}
\toprule
$H$ & Observable & Average warning time\\ \midrule
$[25, 32]$ & 73.5 \% & 0.64 days\\ \midrule
$[25, 30]$ & 89.6 \% & 1.61 days\\
$(30, 32]$ & 72.1 \% & 0.54 days\\ \bottomrule
\end{tabular}
\caption{Relative number of observable impactors and average warning time, for different ranges in absolute magnitude, using two Flyeye telescopes: one at Monte Mufara and one at La Silla.}
\label{tab:obsvsmissed2tel}
\end{table}

When operating two Flyeye telescopes, one at the Northern hemisphere on Monte Mufara and one at the Southern hemisphere in La Silla, it is found that all impacting NEOs approach at least one of the two ground-stations with an elevation above the 15$\degree$ threshold at some point in the trajectory. Approximately 25\% of the impactors exceed the required elevation only at Monte Mufara, and about 23\% have the correct elevation only at La Silla. Then, the remaining 52\% fulfil the elevation condition at both locations at some point in time. This means that without La Silla, 23\% of the impactors will be missed at Monte Mufara for purely geometrical reasons because they approach with a too southern declination.

\subsection{Estimation of Earth impactors approaching from dayside}
For those NEOs that will approach the Earth directly from the Sun, ground-based surveys will not detect them.

\begin{figure}
    \centering
    \includegraphics[width=1.0 \columnwidth]{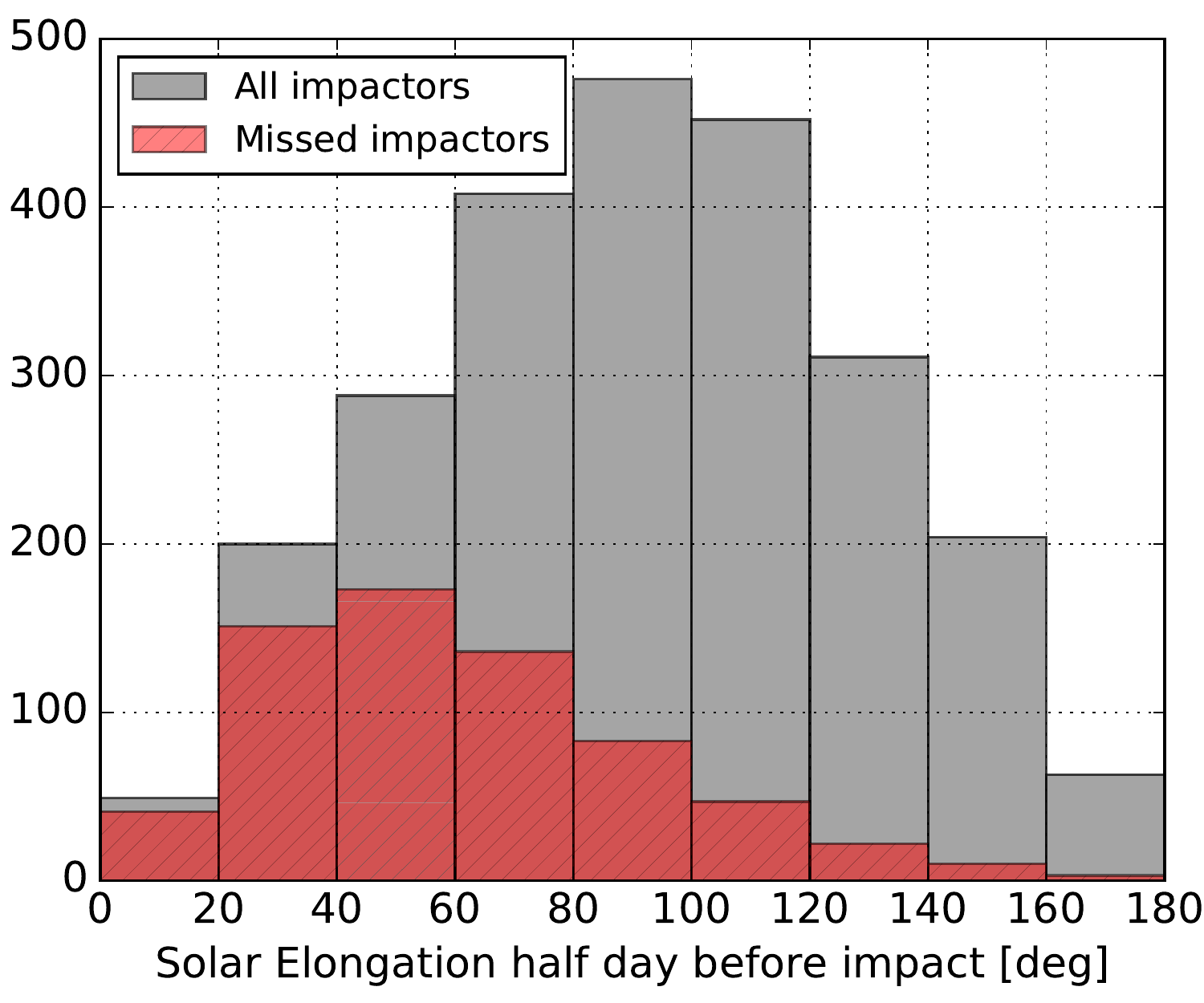}
    \caption{Histogram of the solar elongation computed half day before impact, for the entire impacting NEO population (grey) and the missed impactors (dashed red).}
    \label{fig:histSolarElong}
\end{figure}

We can estimate how many of the missed impactors we are failing to observe, due to the fact that they are arriving from Sun direction, by computing the solar elongation of the object, \textit{i.e.}, the angle Sun-Earth-object. Figure \ref{fig:histSolarElong} shows the solar elongation computed half day before impact of our Earth impactors, which are predominant at solar elongations of 90$\degree$ and few are coming from the Sun.

\begin{figure*}[bp]
    \centering
    \includegraphics[width=1.0 \columnwidth]{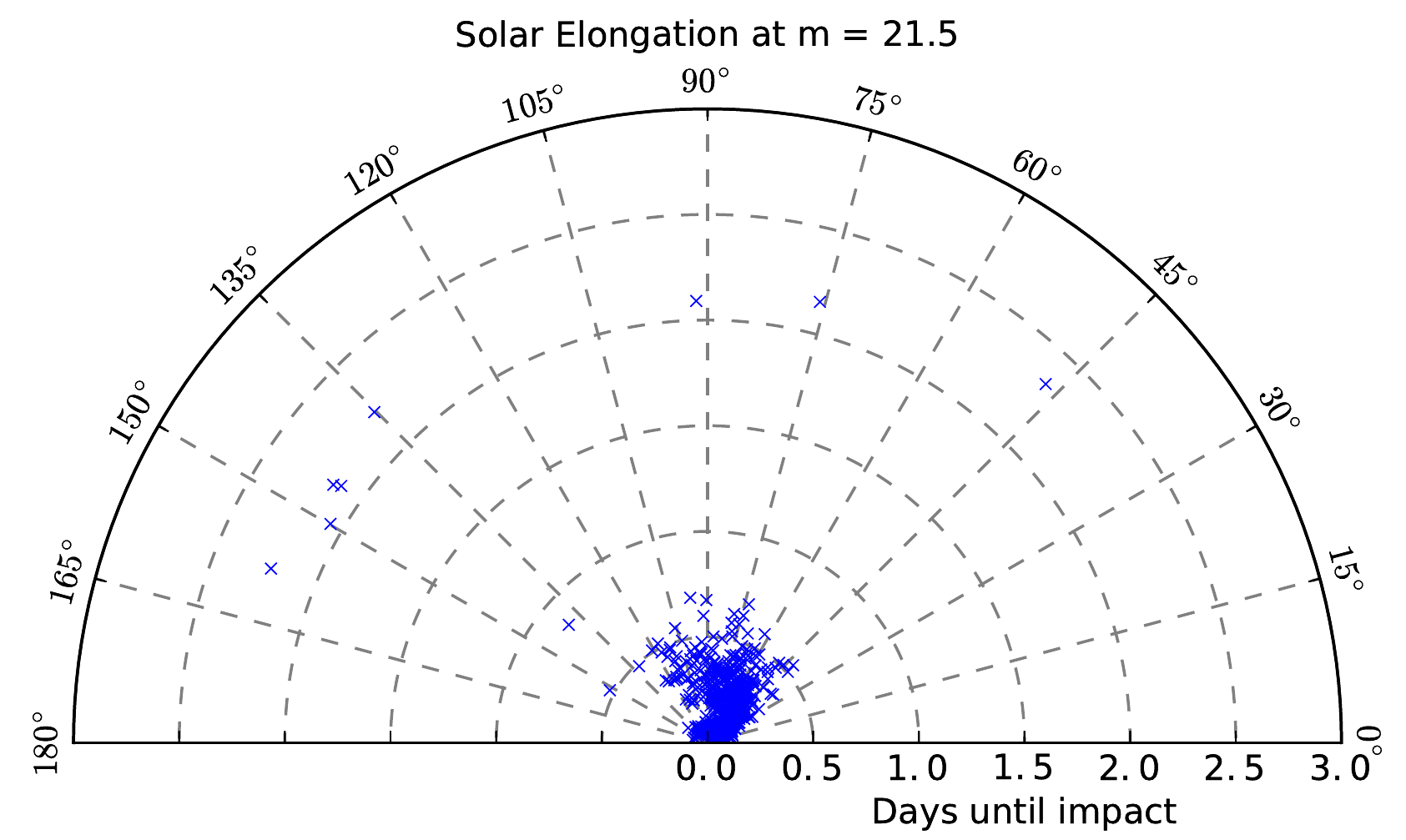}
    \includegraphics[width=1.0 \columnwidth]{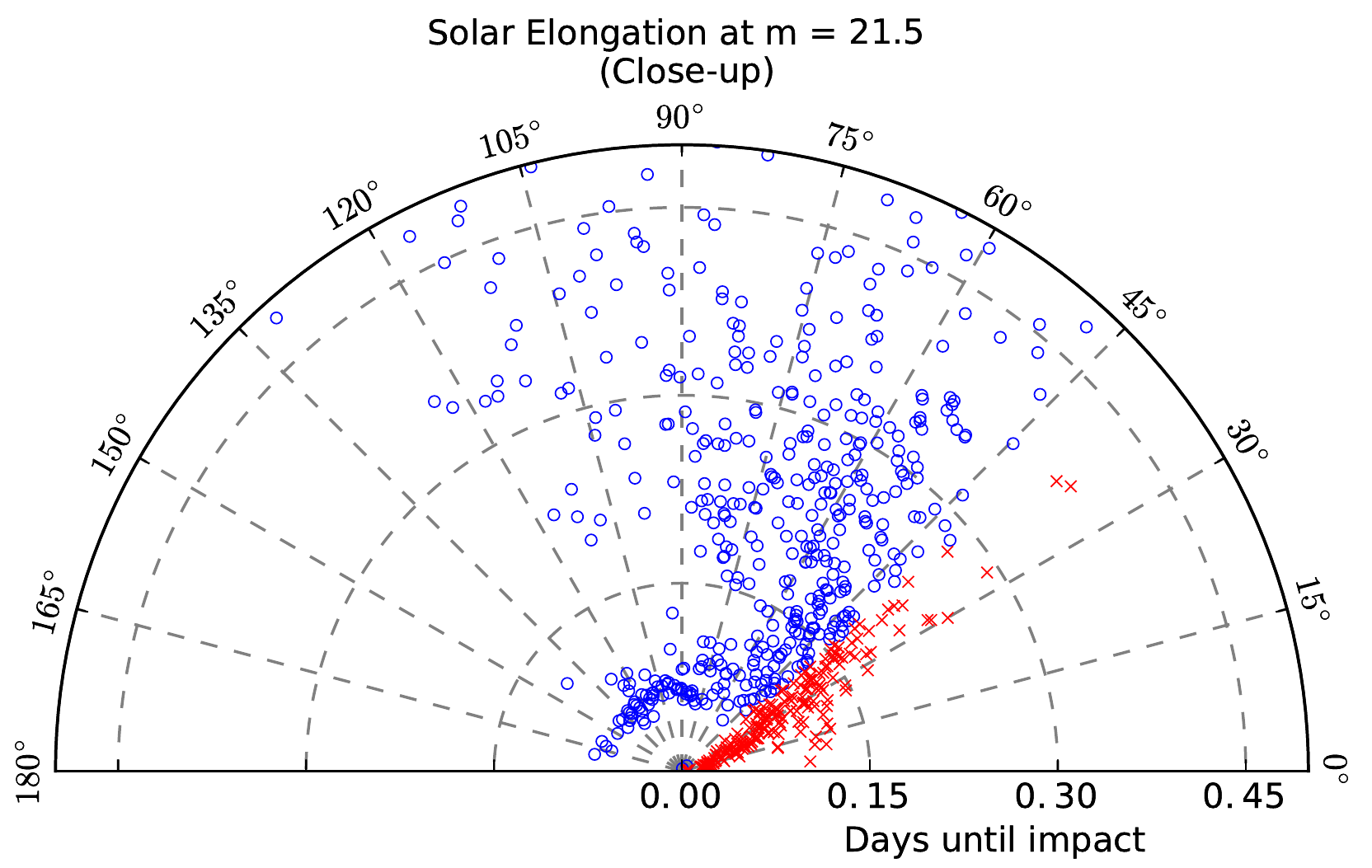}
    \caption{Solar elongation of all missed impactors computed at the time when the apparent magnitude of the object is below 21.5 as a function of the number of days until impact. The plot on the right is a close-up plot of the one on the left, where the \rm{\footnotesize{($\times$)}} {\it represent those objects with a solar elongation below 40$\degree$ and the} \rm{\footnotesize{($\circ$)}} {\it represent those objects with a solar elongation above 40$\degree$.}}
    \label{fig:SolarElong_daystilImp}
\end{figure*}

Figure \ref{fig:SolarElong_daystilImp} shows the solar elongation of all missed impactors from both telescopes considered, computed at the time when the apparent magnitude is below 21.5 as a function of the number of days until impact. As expected, most missed impacting NEOs arrive with low solar elongations or have very short observation windows.

Farnocchia et al.\;(2012) \cite{valsecchi} derived an analytical expression to determine the fraction of impacting NEOs that should be discoverable by ground-based telescopes, based on the angular distance of the impactor from the Sun. Taking a minimum solar elongation of 40$\degree$, as suggested in \cite{valsecchi}, we obtained that about 15.6\% of the Earth impactors from our population would not be detectable with ground-based surveys.

However, the approach by Farnocchia et al.\;(2012) \cite{valsecchi} is simply geometric; the real detection threshold is set by the SNR of the object, \textit{i.e.}, an object with a solar elongation below 40$\degree$ could also be observed if its SNR is high enough. In the simulations, about half of the Earth impactors that always remained below the minimum solar elongation of 40$\degree$ until impact could still be detected because their SNR exceeded the detection threshold and they fulfilled the geometric observation conditions for at least one of the two telescopes. 

We would like to highlight the fact that these estimates are based on our impactor population. When applying the geometric approach by Farnocchia et al.\;(2012) \cite{valsecchi} to the population of 4\,950 synthetic impactors derived by Chesley et al.\;(2004) \cite{chesley}, it is estimated that 19.9\% of the impacting NEOs are approaching from dayside. 

\subsection{Estimation of number of impacts per year detected by the Flyeye telescope}
Although the simulations show that the Flyeye telescope is able to observe about 47\% of the entire impactor population, using only one telescope, the extremely short warning times obtained indicate that a lot of these observable NEOs are likely to be missed because the telescope may not be pointing to the right field in the sky.

As estimated in Section \ref{sec:obspopH}, almost 94\% of the observed impacts have a warning time of less than two days. In order to take into account the fact that for warning times shorter than two days the impactor might be missed, a weight is applied to each observable impactor such that for a warning time of at least two days, the weight is one, but for a shorter warning time, the weight is the warning time divided by two days.

\begin{figure}
    \centering
    \includegraphics[width=1.0 \columnwidth]{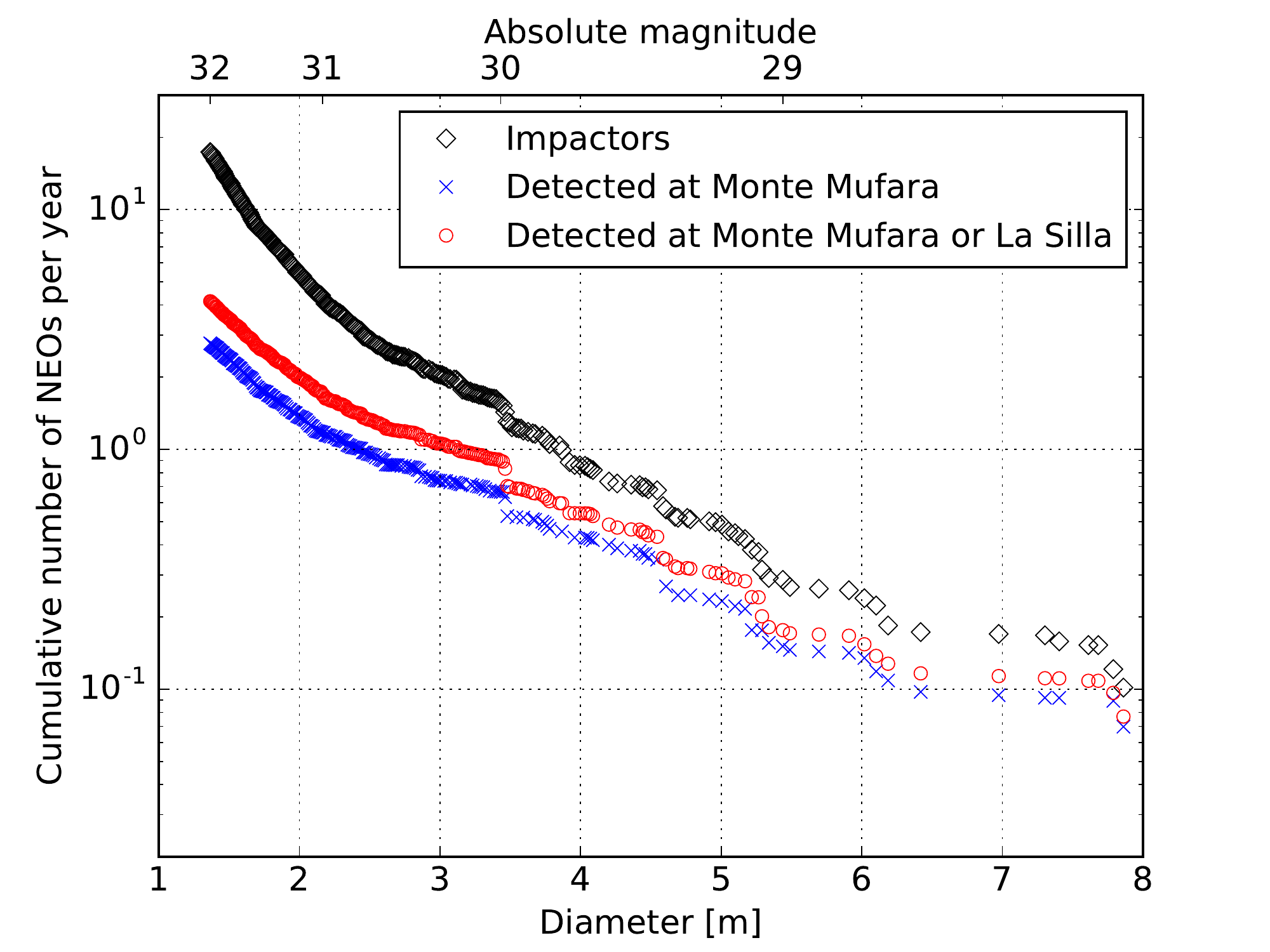}
    \caption{Comparison between the cumulative number of objects colliding with the Earth per year \rm{\footnotesize{($\diamond$)}}{\it, the cumulative number of detected impacts per year using the Flyeye telescope at Monte Mufara} \rm{\footnotesize{($\times$)}} {\it and the cumulative number of detected impacts per year using both Flyeye telescopes at Monte Mufara and La Silla} \rm{\footnotesize{($\circ$)}}.}
    \label{fig:impdetectedperyear}
\end{figure}

Figure \ref{fig:impdetectedperyear} compares the cumulative number of impacts per year that are expected to happen and the cumulative number of impacts per year that we expect to detect when using only the Flyeye telescope at Monte Mufara and when operating both telescopes at Monte Mufara and La Silla.

First, it is estimated that we will observe about 2.8 impacts per year using the Flyeye telescope at Monte Mufara and about 2.9 impacts using the one at La Silla. Considering that we can expect about 17.4 impacts per year from objects down to 1 m, we can conclude that we will only observe about 16--17\% of all impacts with only one telescope.

Second, when we operate both telescopes simultaneously, the number of impacts per year that we can expect to detect increases to 4.1 impacts.

Third, although it cannot yet be seen in Figure \ref{fig:impdetectedperyear}, the total number of detected impacts per year will approach a limit because the Flyeye telescopes will not be able to detect many more impacts even if the simulations are extended to objects fainter than absolute magnitudes of 32; the warning times for such objects will be too short to make a significant statistical contribution.

Nonetheless, it is important to mention that these estimates are obtained assuming a 100\% telescope availability and 365 clear nights per year for both telescopes, which leads to an overestimation of the detection rates.

\section{Conclusions}
A population of synthetic Earth impactors based on the NEO population model developed by Granvik et al.\;(2018) \cite{GRANVIK2018181} was created and validated by comparing the estimates of number of impacts per year to theoretical models and observation data. This impactor population consists of 2\,451 NEOs in the diameter range 1--34 m.

The Flyeye telescope at Monte Mufara is estimated to be able to observe about 61\% of the impactors down to 8 m, about 1.7 days on average. However, for the smallest NEOs in the impactor population in the diameter range 1--3 m, 46\% are observable but with average warning times of half a day, which means that they are likely to be missed.

It is estimated that only 6.4\% of the observed impacting NEOs will have a warning time of two days or more. The main reason for such short warning times is that the objects can only be `seen' when they are very close to the observer, right before impact. This is mainly due to the high absolute magnitudes of the population considered, which result in very faint and small NEOs.

Due to the very short warning times, most of the observable impactors will go undiscovered. When we take into account the warning time of each observed impactor, the estimates show that we can expect to detect almost three impacts per year using one Flyeye telescope. When operating two telescopes simultaneously, one at the Northern hemisphere on Monte Mufara and one at the Southern hemisphere in La Silla, the number of observed impacts per year predicted increases to four.

When simulating two telescopes, the relative number of observable NEOs increases from 47\% to almost 74\%. In addition, our simulations also show that only 52\% of the impactors fulfil the 15$\degree$ elevation condition to be observed at both locations, while 25\% exceed the required elevation only at Monte Mufara and 23\% only at La Silla. This means that adding a second telescope at La Silla not only significantly improves the detection rate of NEOs, but also without this telescope in the Southern hemisphere we would miss completely 23\% of NEOs due to purely geometrical reasons.

Finally, it is estimated that 15.6\% of the Earth impacting NEOs are approaching us from the Sun. Therefore, ground-based telescopes cannot be used if we wish to observe such NEOs and alternatives, such as the use of space-based telescopes, should be studied further. 

\section*{Acknowledgments}
We thank Toni Santana-Ros and Juan L. Cano for their assistance in using ESA's Near-Earth Object Population Observation Program (NEOPOP). 

We thank our colleagues from the SSA ICT team for their excellent hardware and software maintenance services. 

We are very grateful to Marco Micheli and Giovanni B. Valsecchi for their constructive comments and nice suggestions and to Mikael Granvik for providing us with the most updated version of the NEO population model. 

We would also like to show our gratitude to the flight dynamics division at ESA/ESOC for sharing their Mission Analysis Software with us.

\end{document}